%% file: main.tex
 \documentclass[final,5p,times,twocolumn,authoryear]{elsarticle}

\usepackage[colorlinks=true,linkcolor=black, citecolor=blue, urlcolor=blue]{hyperref}

\usepackage{amssymb}
\usepackage{amsmath}
\usepackage{lipsum}

\usepackage{aas_macros}
\usepackage{hyperref}
\usepackage{xspace}
\usepackage{nth}
\usepackage{listings}
\usepackage{xcolor}
\lstdefinestyle{code}{
  basicstyle=\ttfamily\small,
  breaklines=true,           
  breakatwhitespace=false,   
  columns=fullflexible,      
  keepspaces=true,           
  showstringspaces=false,
  tabsize=2,
}

\usepackage{dcolumn}

\definecolor{codegreen}{rgb}{0,0.6,0}
\definecolor{codegray}{rgb}{0.5,0.5,0.5}
\definecolor{codepurple}{rgb}{0.58,0,0.82}
\definecolor{backcolour}{rgb}{0.95,0.95,0.92}
\lstdefinestyle{syntax_highlight}{
  backgroundcolor=\color{backcolour}, commentstyle=\color{codegreen},
  keywordstyle=\color{magenta},
  numberstyle=\tiny\color{codegray},
  stringstyle=\color{codepurple},
  basicstyle=\ttfamily\footnotesize,
  breakatwhitespace=false,         
  breaklines=true,                 
  captionpos=b,                    
  keepspaces=true,                 
  numbers=left,                    
  numbersep=5pt,                  
  showspaces=false,                
  showstringspaces=false,
  showtabs=false,                  
  tabsize=2
}

\usepackage{graphicx}

\usepackage{threeparttable} 
\usepackage{siunitx}
\newcommand{\Msun}{\text{M}_\odot}

\DeclareSIUnit\kpc{kpc}
\DeclareSIUnit\Mpc{Mpc}
\DeclareSIUnit\Gpc{Gpc}
\DeclareSIUnit\Gyr{Gyr}

\newcommand{\codename}[1]{\textcolor{black}{\sc #1}\xspace}        
\newcommand{\simulationname}[1]{\textcolor{black}{\sc #1}\xspace}  

\newcommand{\swift}{\codename{Swift}}
\newcommand{\swiftsimio}{\codename{swiftsimio}}
\newcommand{\hdfstream}{\texttt{hdfstream}\xspace}
\newcommand{\hbt}{\codename{HBT-HERONS}}
\newcommand{\soap}{\codename{SOAP}}

\newcommand{\flamingo}{\simulationname{flamingo}}

\newcommand{\lcdm}{$\Lambda$CDM\xspace}

\journal{Astronomy $\&$ Computing}

\begin{document}

\begin{frontmatter}



\title{The FLAMINGO simulations data release}


\author[icc]{John~C.~Helly}
\affiliation[icc]{organization={Institute for Computational Cosmology, Department of Physics, University of Durham},
            addressline={South Road}, 
            city={Durham},
            postcode={DH1 3LE}, 
            country={UK}}
\author[ul]{Robert~J.~McGibbon}
\affiliation[ul]{organization={Leiden Observatory, Leiden University},
            addressline={PO Box 9513}, 
            city={Leiden},
            postcode={2300 RA}, 
            country={the Netherlands}}
\author[ul]{Joop~Schaye}
\author[ul,ul2]{Matthieu~Schaller}
\affiliation[ul2]{organization={Lorentz Institute for Theoretical Physics, Leiden University},
            addressline={PO Box 9506}, 
            city={Leiden},
            postcode={2300 RA}, 
            country={the Netherlands}}
\author[ul]{William~McDonald}
\author[ul,mpia]{Joey~Braspenning}
\affiliation[mpia]{organization={Max-Planck-Institut f\"ur Astronomie},
            addressline={K\"onigstuhl 17}, 
            city={Heidelberg},
            postcode={D-69117}, 
            country={Germany}}
\author[ul,ul2]{Jeger~C.~Broxterman}
\author[ljmu]{Emily~E.~Costello}
\affiliation[ljmu]{organization={Astrophysics Research Institute, Liverpool John Moores University},
            addressline={ 146 Brownlow Hill}, 
            city={Liverpool},
            postcode={L3 5RF}, 
            country={UK}}
\author[icc]{Willem~Elbers}
\author[ul]{Victor~J.~Forouhar~Moreno}
\author[icc]{Carlos~S.~Frenk}
\author[icc]{Adrian~Jenkins}
\author[ul]{Roi~Kugel}
\author[ljmu]{Ian~G.~McCarthy}
\author[ljmu]{Jaime~Salcido}
\author[ul]{Marcel~P.~van~Daalen}
\author[ul]{Bert~Vandenbroucke}
\author[ljmu]{Tianyi~Yang}

\begin{abstract}
We describe the public release of $>2.3$~petabytes of data from the \flamingo cosmological simulations. The suite consists of hydrodynamical simulations that include radiative cooling, star formation, stellar mass loss and the resulting chemical enrichment, supernova feedback, and two implementations of AGN feedback. Neutrinos are simulated explicitly using particles. Data products include snapshots, halo and galaxy catalogues, HEALPix all-sky lightcone maps, particle data for lightcone maps, and power spectra. The \flamingo set includes 22 hydrodynamical simulations. In addition, there are 16 gravity-only simulations, including the $10\,080^3$ particles \flamingo-10k run, with initial conditions that match those of the corresponding hydrodynamical runs. The fiducial hydrodynamical simulations span three numerical resolutions that have each been calibrated to reproduce the present-day galaxy stellar mass function and gas fractions in low-redshift clusters. Other simulations systematically vary the galaxy stellar mass function, cluster gas fractions, cosmology (including neutrino masses), and/or the nature of dark matter, in volumes of 1~cGpc$^3$. The release includes hitherto unpublished simulations that use extra dark matter particles. 
While we provide a facility for downloading complete simulation outputs, we recognise that for many users this will not be possible due to limited local storage or network bandwidth. We implement a web service that enables users to explore available outputs and selectively download datasets or parts of datasets.
\end{abstract}




\end{frontmatter}




\input{introduction}

\input{flamingo_simulations}

\input{data_products}

\input{data_access}

\input{discussion}

\input{conclusions}

\section*{Acknowledgements}
We are grateful to all FLAMINGO users who helped identify issues with the data and documentation.

JCH is supported by STFC consolidated grant ST/X001075/1. This project has received funding from the Netherlands Organization for Scientific Research (NWO) through research programme Athena 184.034.002. VJFM acknowledges support by NWO through the Dark Universe Science Collaboration (OCENW.XL21.XL21.025). This work was supported by the Science and Technology Facilities Council (grant number ST/Y002733/1). This project has received funding from the European Research Council (ERC) under the European Union’s Horizon 2020 research and innovation programme (grant agreement No 769130). This work used the DiRAC@Durham facility managed by the Institute for Computational Cosmology on behalf of the STFC DiRAC HPC Facility (\url{www.dirac.ac.uk}). The equipment was funded by BEIS capital funding via STFC capital grants ST/K00042X/1, ST/P002293/1, ST/R002371/1 and ST/S002502/1, Durham University and STFC operations grant ST/R000832/1. DiRAC is part of the National e-Infrastructure.

For the purpose of open access, the authors have applied a Creative Commons attribution (CC BY) licence to any Author Accepted Manuscript version arising.

\section*{Data access statement}

The \flamingo simulation outputs described in this paper are available on the web at \url{https://flamingo.strw.leidenuniv.nl/data.html}.

\appendix

\input{examples}


\bibliographystyle{elsarticle-harv} 
\bibliography{references,main}






\end{document}

%% file: introduction.tex
\section{Introduction}
\label{introduction}
Cosmological simulations are widely used to study a large variety of topics in galaxy formation and cosmology. Simulations are needed to predict observational quantities, such as the matter power spectrum or the halo mass function, enabling the measurement of the cosmological parameters on which they depend. Simulations are also required to model factors that complicate the analysis of observations, such as selection and projection effects. 

Dark-matter-only (DMO) simulations, i.e.\ simulations that account for the effects of baryons in the initial conditions but not for the subsequent evolution, have long been the workhorse for comparisons with observational surveys. However, while baryonic matter is expected to trace the dark matter on very large scales, on non-linear scales ($\lesssim 10$~Mpc), cooling flows, star formation, and galactic winds driven by stellar and active galactic nucleus (AGN) feedback can alter the distribution of baryons. As observational methods become more precise and probe smaller scales, baryonic effects become increasingly important. Indeed, modelling baryonic effects is, for example, essential for ongoing weak lensing surveys such as \textit{Euclid} \citep[e.g.][]{Semboloni2011} and \textit{LSST} \citep[e.g.][]{Robertson2026}. 

While baryonic effects can be modelled by post-processing DMO simulations \citep[e.g.][]{Schneider2015,Arico2021emulator}, or by using simplified models built on top of DMO frameworks \citep[e.g.][]{Mead2015,Semboloni2013,Debackere2020}, such models rely on hydrodynamical simulations to motivate, optimize and validate their methods. 

Furthermore, observations rely on the detection of baryonic matter, either diffuse gas or stars. Examples of the former are secondary cosmic microwave background (CMB) anisotropies, such as the thermal and kinetic Sunyaev-Zel'dovich (SZ) effects, X-ray emission, and fast radio burst (FRB) dispersion measures. Examples of the latter are galaxy clustering and redshift space distortions. In the case of gas, the signal depends on the physical conditions such as the temperature, density, and chemical composition, while in the case of stars it depends on the way in which galaxies populate dark matter haloes. Hydrodynamical simulations can directly predict such observables.  

A drawback of using hydrodynamical simulations, as opposed to post-processed DMO simulations, is that hydrodynamical simulations are more expensive computationally, both in terms of computing time and memory/storage requirements. In small volumes (side length $L \lesssim 10^2\,$Mpc), which can be simulated at relatively high resolution (particle mass $\lesssim 10^6\,$M$_\odot$), it is possible to model the low-mass end of the galaxy population and the detailed properties of massive galaxies. Examples of such simulations are HorizonAGN \citep{Dubois2014}, EAGLE \citep{Schaye2015,Crain2015}, IllustrisTNG \citep{Pillepich2018TNG,Nelson2019}, Simba \citep{Dave2019}, ASTRID \citep{Bird2022}, FIREbox \citep{Feldmann2023}, CROCODILE \citep{Oku2024}, (X)FABLE \citep{Bigwood2025a}, and COLIBRE \citep{Schaye2025,Chaikin2025calibration}. However, these simulations do not provide good statistics for rare objects, such as galaxy clusters, and they cannot model the distribution of matter on scales comparable to or larger than their box size. Hence, for cosmology and the study of clusters and massive galaxies, large volumes ($L \gtrsim 10^3\,$Mpc), which can currently only be modelled with relatively low resolution ($\gg 10^6\,$M$_\odot$), tend to be required. Examples of such simulations are Cosmo-OWLS \citep{LeBrun2014}, BAHAMAS \citep{McCarthy2017}, MillenniumTNG \citep{Pakmor2023}, Magneticum Pathfinder \citep{Dolag2025}, and Frontier-E \citep{Frontiere2025}. 

Another drawback of hydrodynamical simulations is that they depend on uncertain subgrid processes. To maximize the utility of hydrodynamical simulations, it is therefore important to calibrate them to observations that are relevant to the goals of the project and to vary the uncertain ingredients. This was the approach taken for the BAHAMAS simulations, for which the feedback was varied relative to a model that reproduces the galaxy stellar mass function, which is important for the galaxy-halo connection, and cluster gas fractions, which is important for the matter power spectrum. In addition, to study the interplay between variations in cosmology and baryonic physics, it is necessary to vary the cosmology. 

Here we present the public release of the data from the Virgo Consortium's \flamingo\footnote{The acronym stands for \textbf{F}ull-hydro \textbf{L}arge-scale structure simulations with \textbf{A}ll-sky \textbf{M}apping for the \textbf{I}nterpretation of \textbf{N}ext \textbf{G}eneration \textbf{O}bservations.} project \citep{Schaye2023,Kugel2023}. \flamingo consists of a large suite of cosmological hydrodynamical simulations and their DMO counterparts. Compared with simulations focusing on galaxy formation, the volumes are large ($L\ge1$~cGpc), but the resolution is modest (baryonic particle mass $\ge 1\times10^8\,$M$_\odot$). The simulations include not only cold dark matter and baryons, but also neutrinos. The subgrid physics in the fiducial model is calibrated on the observed $z=0$ galaxy stellar mass function, gas fractions of low-$z$ clusters, and black hole masses in massive present-day galaxies. The calibration is repeated for three different numerical resolutions, spanning two orders of magnitude in particle mass. The suite currently includes ten variations in baryon physics, which systematically vary the galaxy stellar mass function and/or cluster gas fractions. It also includes two different implementations of AGN feedback, thermally-driven winds or kinetic jets. In addition, there are nine different cosmologies (ten for gravity-only), including variations in the parameters of the $\Lambda$CDM model, the neutrino mass, decaying dark matter, and a hitherto unpublished cosmology assuming evolving dark energy (the last model is currently only available as DMO).

Following the pioneering work of \cite{MillenniumDataRelease}, who presented a publicly-available relational database containing the halo and galaxy properties of the Millennium simulation \citep{2005Natur.435..629S}, partial or full releases of the large datasets generated by cosmological simulation campaigns have become a common practice. This has enabled a very large number of studies by scientists not directly affiliated with the teams producing the simulations and has been a major step for open science. The practice was later extended beyond semi-analytic models with examples including the Illustris full data release \citep{IllustrisDataRelease}, the EAGLE galaxy catalogs and merger trees \citep{EAGLEDataRelease1} followed by the EAGLE snapshot data \citep{EAGLEDataRelease2}, and the release of the IllustrisTNG suite (TNG50, TNG100, and TNG300; \cite{TNGDataRelease}). More recently, there have been public releases of zoom-in simulations of suites of individual haloes from FIRE-2 \citep{FIRE2DataRelease}, AURIGA \citep{Grand2024} and TNG-Cluster \citep{TNGCluster}, collections of small-volume simulations \citep[e.g.\ the CAMELS set][which comprises nearly 17\,000 simulations and more than 2~PB of data]{CAMELSDataRelease}, and radiation-magnetohydrodynamical simulations of the high redshift universe including THESAN \citep{THESANDataRelease} and Sphinx20 \citep{SPHINX20DataRelease}.

All of these simulation data releases, with the exception of Sphinx20, provide full downloads of raw simulation data. In some cases, facilities for selective extraction, online search and online analysis are also provided. For example, the Illustris, IllustrisTNG and TNG-Cluster releases provide a web-based API to fetch information about specific objects or cut out parts of snapshots, the EAGLE data release includes a database of halos and galaxies which can be queried with SQL, and CAMELS allows data access via Jupyter notebooks.

We follow these practices and provide both full downloads and a mechanism for selective access to outputs, releasing all the \flamingo data products that we can afford to keep on disc. This includes snapshots, halo and galaxy catalogues, HEALPIX lightcone maps, particle lightcones, and power spectra. The total volume of the data released exceeds 2.4~PB. The data are hosted at the DiRAC Memory Intensive Service\footnote{\url{https://dirac.ac.uk/memory-intensive-durham/}} in Durham, UK, where the simulations were run. The large size of the simulations makes full downloads problematic because many users will not have access to sufficient network bandwidth or storage. We address this by providing a mechanism to request single datasets or parts of datasets from the output files. Users with more limited computing resources may still be able to exploit subsets of the data, such as particular galaxies extracted from the snapshots or halo catalogues containing just a few quantities of interest. For ease of use, this system is designed to resemble the tools researchers would use to work with local data.

Since the publication of \citet{Schaye2023}, \flamingo has already resulted in $\approx 80$ refereed papers\footnote{See \url{https://flamingo.strw.leidenuniv.nl/papers.html}.} on topics such as high-redshift galaxies, clusters, and large-scale structure. Some of these papers have resulted in additional data products that are also part of this release. Thanks to the efforts of the authors of completed and upcoming publications, the strengths and weaknesses of the simulations are fairly well characterized and documented, but we have only scratched the surface of the potential scientific applications. 

For (interactive) visualisations and videos of the \flamingo simulations, see \url{https://flamingo.strw.leidenuniv.nl/}. For access to the data and online documentation, see \url{https://flamingo.strw.leidenuniv.nl/data.html}.

This paper is organized as follows. The \flamingo simulations and data products are described in Sections~\ref{sec:simulations} and \ref{sec:data_products}, respectively. Section~\ref{sec:data_access} discusses the design and implementation of the data service. In Section~\ref{sec:discussion} we briefly summarize the results of published comparisons with data, the interpretation of comparisons between the different \flamingo resolutions, and known issues and bugs. Section~\ref{sec:credits} provides advice on how to acknowledge the usage of the released data. Finally, we conclude in Section~\ref{sec:conclusions}.

%% file: flamingo_simulations.tex
\section{FLAMINGO simulations} \label{sec:simulations}
This section contains a brief overview of the \flamingo\ simulations. Further details on the simulation methods and the galaxy formation model are provided in \citet{Schaye2023}, while the calibration of the subgrid model is detailed in \citet{Kugel2023}.

\subsection{The FLAMINGO model}

All simulations were performed using the open-source simulation code \swift
\citep{SWIFT}\footnote{Publicly available, including the exact version used for
the \flamingo runs, at \url{www.swiftsim.com}. The parameter files for the various \flamingo simulations are provided on the data release website as well as in the \swift code repository. }. In particular, gravity is solved for using a \nth{4}-order fast-multipole method \citep[e.g.][]{Cheng1999} coupled to a particle-mesh method for long-range forces \citep[e.g.][]{Bagla2003}, the gas is evolved using the SPHENIX \citep{Borrow2022} flavour of Smoothed Particle Hydrodynamics (SPH), and neutrinos are followed using the $\delta f$-method
of \cite{Elbers2021}.

The  \flamingo simulations include subgrid prescriptions for radiative cooling and 
heating following \cite{Ploeckinger2020}, star formation and an entropy floor for the dense gas
based on the method of \cite{Schaye2008}, the stellar evolution and chemical enrichment model of
\cite{Wiersma2009}, and core collapse supernova feedback implemented kinetically
following \cite{Chaikin2023}. Supermassive black holes seeding, gas accretion, merging, and 
repositioning are modelled using ingredients from \cite{Springel2005}, \cite{Booth2009} and \cite{Bahe2022}. The feedback from AGN is modelled either as thermally-driven winds \citep{Booth2009} or using the
collimated jet model of \cite{Husko2022}.

\flamingo includes three different resolutions, labelled by the rounded log base 10 of the baryonic particle mass in units of the solar mass: the high-resolution m8, intermediate-resolution m9, and low-resolution m10 (see Table~\ref{tab:simulations}). The variations in cosmology and/or galaxy formation physics all use m9 resolution in 1~cGpc boxes. The fiducial model is available at all resolutions and, at m9 resolution, also in a larger volume of 2.8~cGpc on a side. 

As described by \cite{Kugel2023}, four free parameters of the subgrid models associated with stellar and AGN feedback were calibrated using a Gaussian process emulator, trained on a Latin hypercube of simulations of small volumes, to predict the observables as a function of the input parameters. A Monte Carlo Markov Chain search was then used in combination with the emulator to find the best-fitting value reproducing the target
datasets, while accounting for observational errors. In particular, the fiducial model was calibrated to match the $z=0$ galaxy stellar mass function from the DR4 release of the GAMA survey \citep{Driver2022} and
simultaneously the gas fractions in low-redshift clusters inferred from (pre-eRosita) X-ray and weak-lensing data aggregated by
\cite{Kugel2023}. For the galaxy mass function the lower limit of the mass range used in the calibration depends on the simulation resolution: $M_* > 10^{8.67}$, $10^{9.92}$, and $10^{11.17}\,\text{M}_\odot$ for m8, m9, and m10, respectively, while the upper limit is always $10^{11.5}\,\text{M}_\odot$. For the cluster gas fractions, the lower limit of the halo mass considered during the calibration is always $M_\text{500c} = 10^{13.5}\,\text{M}_\odot$, while the upper mass limits are $10^{13.73}$, $10^{14.36}$, and $10^{14.53}\,\text{M}_\odot$ for m8, m9, and m10, respectively. 

Besides generating sets of simulation parameters matching the
data, additional simulations were generated where the target data was shifted by
particular amounts with respect to observations. For the cluster
gas fraction variations, the observed gas fractions were shifted up and down compared to the results by $\pm N\sigma$, where $\sigma$ is the error in the data \citep[see][for the exact
definitions]{Kugel2023}. Similar variations were generated for shifts in the stellar
mass function. The free parameters were calibrated independently for simulations 
using the thermal and jet AGN models and for simulations using different resolutions.

Since the $z = 0$ galaxy stellar mass function and the gas fractions of clusters in the mass and redshift ranges above were directly used to calibrate the \flamingo model parameters, these quantities (and closely related quantities such as the stellar mass to halo mass relation and cluster stellar and baryon mass fractions in the corresponding mass ranges) should not be treated as predictions of the model. Additionally, \flamingo uses the same AGN feedback efficiency parameters as \cite{Booth2009}, who chose values which reproduce the normalisation of the relation between galaxy stellar mass and central black hole mass. This relation is therefore a product of a previous calibration. Other halo and galaxy properties, such as those presented in section~6 of \cite{Schaye2023}, were not used in the calibration process and may be considered to be predictions of the model.

The initial conditions (ICs) were generated using the \codename{MonofonIC} code
\citep{Hahn2021, Elbers2022} using a 3-fluid formalism with a separate transfer
function for each of the species. The baseline cosmological model used for the simulations is the 
maximum likelihood values of the flat \lcdm + neutrinos model fitted to the DES year 3 data release
\citep{Abbott2022} combined with external probes, i.e. their ``3$\times$2pt +
All Ext.''  model. The parameter values are given in the first row of Table \ref{tab:cosmologies}.

\subsection{Model variations}

The main properties of the 22 hydrodynamical and 16 DMO simulations released here\footnote{This set extends the original 16 hydrodynamical and 12 DMO runs presented by \cite{Schaye2023} with additional model variations, primarily varying the values of the cosmological parameters.} are given in Tables~\ref{tab:simulations} and \ref{tab:DMO_simulations}, respectively. Most runs simulate a volume of $(1~{\rm cGpc})^3$ with the intermediate resolution (`m9') model, which corresponds to a (initial) mean baryonic particle mass of $1.07\times10^9~\Msun$ and a mean cold dark matter particle mass of $5.65\times10^9~\Msun$. The cosmological variations (with parameters given in Table~\ref{tab:cosmologies}) are all performed at this resolution, and so are the feedback variations. The set of m9 runs also includes the `L2p8\_m9' simulation, which uses the fiducial feedback and cosmology but in a volume of $(2.8~{\rm cGpc})^3$, i.e.\ in a volume 22x larger, and was the largest cosmological hydrodynamical simulation (by number of resolution elements) run to $z=0$ when first published. The two simulations `L1\_m8' and `L1\_m10' are run in the same $(1~{\rm cGpc})^3$ volume as the L1\_m9 simulation, but use 8x higher and lower mass resolutions, respectively, with the subgrid parameter calibration performed at each resolution as described above. Apart from the simulation L1\_m9\_extraDM, all runs use equal numbers of baryonic and dark matter particles, while the number of neutrino particles is a factor $1.8^3$ smaller. All simulations in a given volume use the same phases in the initial conditions, allowing objects to be matched between models.

The cosmology variations (see Table~\ref{tab:cosmologies}) include runs based on the best-fitting \lcdm model to the Planck data with larger neutrino masses, the low-$\sigma_8$ `lensing cosmology' of \cite{Amon2023}, and \lcdm extensions allowing for decaying dark matter \cite[See][for details]{Elbers2024}.

In addition to the feedback variations described above, there are two hydrodynamical simulations at m9 resolution in the 1~cGpc box. In the `NoCooling' run radiative cooling (but not photo-heating), star formation, and feedback have all been switched off. Besides the equations of gravity and hydrodynamics, only heating sources (by adiabatic compression, shocks, and photoionization by the UV background radiation) are included in the physics. The last variation is the `L1\_m9\_extraDM' model, which uses
the fiducial calibration but super-samples the dark matter by a factor of 4. This leads to baryon and dark matter particles with similar masses and thus reduces the spurious energy transfer between species highlighted by \cite{Ludlow2019,Ludlow2021,Ludlow2023}. 

DMO counterparts to all the runs above (except for `L1\_m9\_extraDM') are also available. Accompanying these, we also provide the \flamingo-10k simulation, which was first used in \citet{Pizzati2024}. It features the $(2.8~{\rm cGpc})^3$ volume but at `m8' resolution. This simulation evolves nearly $1.2\times 10^{12}$ particles. The semi-analytic model \textsc{Galform} has been run on the output and will soon be available as an additional high-resolution galaxy catalogue.

\begin{table*}
\footnotesize
	\centering
	\caption{Hydrodynamical simulations in the \flamingo suite. Table replicated from \citet{Schaye2023} and expanded with newer models. The first four lines list the simulations that use the fiducial galaxy formation model and assume the fiducial cosmology, but use different volumes and resolutions. The remaining lines list the model variations, which all use a 1~cGpc box and intermediate resolution. The columns list the simulation identifier (where m8, m9 and m10 indicate $\log_{10}$ of the mean baryonic particle mass and correspond to high, intermediate, and low resolution, respectively; absence of this part implies m9 resolution); the number of standard deviations by which the observed stellar masses are shifted before calibration, $\Delta m_\ast$; the number of standard deviations by which the observed cluster gas fractions are shifted before calibration, $\Delta f_\text{gas}$; the AGN feedback implementation (thermal or jets); the comoving box side length, $L$; the number of baryonic particles, $N_\text{b}$ (which equals the number of CDM particles, $N_\text{CDM}$, except for the L1\_m9\_extraDM simulation where $N_\text{CDM} = 4\times N_\text{b}$); the number of neutrino particles, $N_\nu$; the initial mean baryonic particle mass, $m_\text{g}$; the mean CDM particle mass, $m_\text{CDM}$; the Plummer-equivalent comoving gravitational softening length, $\epsilon_\text{com}$; the maximum proper gravitational softening length, $\epsilon_\text{prop}$; the assumed cosmology which is specified in Table~\ref{tab:cosmologies}; and the paper reference introducing the simulations: (a) \citet{Schaye2023}, (b) \citet{McCarthy2025}, (c) \citet{Elbers2024}, (d) this paper.}
	\label{tab:simulations}
	\begin{tabular}{lrrlrrrllrrll} 
		\hline
		Identifier & $\Delta m_\ast$ & $\Delta f_\text{gas}$ & AGN & $L$ & $N_\text{b}$ & $N_\nu$ & $m_\text{g}$ & $m_\text{CDM}$ & $\epsilon_\text{com}$ & $\epsilon_\text{prop}$ & Cosmology & Ref. \\ 
		           & ($\sigma$) & ($\sigma$) && (cGpc) &&& ($\Msun$) & ($\Msun$)  & (ckpc) & (pkpc) & \\
		\hline
		L1\_m8             & 0 & 0 & thermal & 1.0 & $3600^3$ & $2000^3$ & $1.34\times 10^8$ & $7.06\times 10^8$    & 11.2  & 2.85 & D3A & a\\
		L1\_m9             & 0 & 0 & thermal & 1.0 & $1800^3$ & $1000^3$ & $1.07\times 10^9$ & $5.65\times 10^9$    & 22.3  & 5.70 & D3A & a\\
		L1\_m10              & 0 & 0 & thermal & 1.0 & $900^3$ & $500^3$ & $8.56\times 10^9$ & $4.52\times 10^{10}$ & 44.6 & 11.40 & D3A & a\\
		L2p8\_m9           & 0 & 0 & thermal &2.8& $5040^3$ & $2800^3$ & $1.07\times 10^9$ & $5.65\times 10^9$    & 22.3  & 5.70 & D3A & a\\
		fgas$+2\sigma$  & 0 & $+2$ & thermal & 1.0 & $1800^3$ & $1000^3$ & $1.07\times 10^9$ & $5.65\times 10^9$ & 22.3  & 5.70 & D3A & a\\
		fgas$-2\sigma$  & 0 & $-2$ & thermal & 1.0 & $1800^3$ & $1000^3$ & $1.07\times 10^9$ & $5.65\times 10^9$    & 22.3  & 5.70 & D3A & a\\
		fgas$-4\sigma$  & 0 & $-4$ & thermal & 1.0 & $1800^3$ & $1000^3$ & $1.07\times 10^9$ & $5.65\times 10^9$    & 22.3  & 5.70 & D3A & a\\
		fgas$-8\sigma$  & 0 & $-8$ & thermal & 1.0 & $1800^3$ & $1000^3$ & $1.07\times 10^9$ & $5.65\times 10^9$    & 22.3  & 5.70 & D3A & a\\
		M*$-\sigma$ & $-1$ & 0 & thermal & 1.0 & $1800^3$ & $1000^3$ & $1.07\times 10^9$ & $5.65\times 10^9$    & 22.3  & 5.70 & D3A & a\\
		M*$-\sigma$\_fgas$-4\sigma$     & $-1$ & $-4$ & thermal & 1.0 & $1800^3$ & $1000^3$ & $1.07\times 10^9$ & $5.65\times 10^9$    & 22.3  & 5.70 & D3A & a \\
		Jet            & 0 & 0 & jets & 1.0 & $1800^3$ & $1000^3$ & $1.07\times 10^9$ & $5.65\times 10^9$    & 22.3  & 5.70 & D3A & a \\
		Jet\_fgas$-4\sigma$  & 0 & $-4$ & jets & 1.0 & $1800^3$ & $1000^3$ & $1.07\times 10^9$ & $5.65\times 10^9$    & 22.3  & 5.70 & D3A & a \\
        NoCooling & n/a & n/a & n/a & 1.0 & $1800^3$ & $1000^3$ & $1.07\times 10^9$ & $5.65\times 10^9$    & 22.3  & 5.70 & D3A & b\\
        L1\_m9\_extraDM & 0 & 0 & thermal & 1.0 & $1800^3$ & $1000^3$ & $1.07\times 10^9$ & $1.41\times 10^9$    & 14.1  & 5.70 & D3A & d\\
		Planck          & 0 & 0 & thermal & 1.0 & $1800^3$ & $1000^3$  & $1.07\times 10^9$ & $5.72\times 10^9$    & 22.3  & 5.70 & Planck& a\\
		PlanckNu0p24Var & 0 & 0 & thermal & 1.0 & $1800^3$ & $1000^3$ & $1.06\times 10^9$ & $5.67\times 10^9$    & 22.3  & 5.70 & PlanckNu0p24Var & a \\
		PlanckNu0p24Fix & 0 & 0 & thermal & 1.0 & $1800^3$ & $1000^3$ & $1.07\times 10^9$ & $5.62\times 10^9$    & 22.3  & 5.70 & PlanckNu0p24Fix & a\\
        PlanckNu0p48Fix & 0 & 0 & thermal & 1.0 & $1800^3$ & $1000^3$ & $1.07\times 10^9$ & $5.62\times 10^9$    & 22.3  & 5.70 & PlanckNu0p48Fix & c\\
		LS8             & 0 & 0 & thermal & 1.0 & $1800^3$ & $1000^3$ & $1.07\times 10^9$ & $5.65\times 10^9$    & 22.3  & 5.70 & LS8 & a\\
        LS8\_fgas$-8\sigma$ & 0 & -8 & thermal & 1.0 & $1800^3$ & $1000^3$ & $1.07\times 10^9$ & $5.65\times 10^9$    & 22.3  & 5.70 & LS8 & b \\
		PlanckDCDM12          & 0 & 0 & thermal & 1.0 & $1800^3$ & $1000^3$  & $1.07\times 10^9$ & $5.72\times 10^9$    & 22.3  & 5.70 & PlanckDCDM12 & c\\ 
        PlanckDCDM24          & 0 & 0 & thermal & 1.0 & $1800^3$ & $1000^3$  & $1.07\times 10^9$ & $5.72\times 10^9$    & 22.3  & 5.70 & PlanckDCDM24 & c\\
		\hline
	\end{tabular}
\end{table*}

\begin{table*}
\footnotesize
	\centering
	\caption{Gravity-only simulations in the \flamingo suite. Table replicated from \citet{Schaye2023} and expanded with newer models. The columns list the simulation identifier; the comoving box side length, $L$; the number of CDM particles, $N_\text{CDM}$; the number of neutrino particles, $N_\nu$; the mean CDM particle mass, $m_\text{CDM}$; the Plummer-equivalent comoving gravitational softening length, $\epsilon_\text{com}$; the maximum proper gravitational softening length, $\epsilon_\text{prop}$; the assumed cosmology which is specified in Table~\ref{tab:cosmologies}; and the paper reference introducing the simulations: (a) \citet{Schaye2023}, (c) \citet{Elbers2024}, (d) this paper, (e) \citet{Pizzati2024}. Simulation L1\_m9\_ip\_DMO is identical to  L1\_m9\_DMO except that the phases in the initial conditions were inverted. Note that there are no hydrodynamical counterparts to FLAMINGO-10k, L5p6\_m10\_DMO, L11p2\_m11\_DMO, PlanckNu0p12Var\_DMO, and L1\_m9\_ip\_DMO.}
	\label{tab:DMO_simulations}
	\begin{tabular}{lrrrlrrll} 
		\hline
		Identifier & $L$ & $N_\text{CDM}$ & $N_\nu$ & $m_\text{CDM}$ & $\epsilon_\text{com}$ & $\epsilon_\text{prop}$ & Cosmology & Ref. \\ 
		           & (cGpc) &&& ($\Msun$)  & (ckpc) & (pkpc) & \\
		\hline

		L1\_m8\_DMO   & 1.0 & $3600^3$ & $2000^3$ & $8.40\times 10^8$    & 11.2  & 2.85 & D3A & a\\
		L1\_m9\_DMO   & 1.0 & $1800^3$ & $1000^3$ & $6.72\times 10^9$    & 22.3  & 5.70 & D3A & a \\
		L1\_m10\_DMO  & 1.0 & $900^3$ & $500^3$ & $5.38\times 10^{10}$ & 44.6 & 11.40 & D3A & a\\
        FLAMINGO-10k & 2.8 & $10080^3$ & $5600^3$ & $8.40\times 10^8$    & 11.2  & 2.85 & D3A & e\\        
		L2p8\_m9\_DMO & 2.8& $5040^3$ & $2800^3$ & $6.72\times 10^9$    & 22.3  & 5.70 & D3A & a\\		
		L5p6\_m10\_DMO & 5.6& $5040^3$ & $2800^3$ & $5.38\times 10^{10}$    & 44.6  & 11.40 & D3A & a\\		
		L11p2\_m11\_DMO & 11.2& $5040^3$ & $2800^3$ & $4.30\times 10^{11}$    & 89.2  & 22.80 & D3A & a\\	
        L1\_m9\_ip\_DMO  & 1.0 & $1800^3$ & $1000^3$ & $6.72\times 10^9$ & 22.3  & 5.70 & D3A & d\\
        Planck\_DMO & 1.0 & $1800^3$ & $1000^3$ & $6.78 \times 10^9$ & 22.3  & 5.70 & Planck & a\\
		PlanckNu0p12Var\_DMO & 1.0 & $1800^3$ & $1000^3$ & $6.74\times 10^9$ & 22.3  & 5.70 & PlanckNu0p12Var & a\\
		PlanckNu0p24Var\_DMO & 1.0 & $1800^3$ & $1000^3$ & $6.73\times 10^9$ & 22.3  & 5.70 & PlanckNu0p24Var & a\\
		PlanckNu0p24Fix\_DMO & 1.0 & $1800^3$ & $1000^3$ & $6.68\times 10^9$ & 22.3  & 5.70 & PlanckNu0p24Fix & a\\
    	PlanckNu0p48Fix\_DMO & 1.0 & $1800^3$ & $1000^3$ & $6.68\times 10^9$ & 22.3  & 5.70 & PlanckNu0p48Fix & c\\
		LS8\_DMO & 1.0 & $1800^3$ & $1000^3$ & $6.72\times 10^9$ & 22.3  & 5.70 & LS8 & a\\
        PlanckDCDM12\_DMO & 1.0 & $1800^3$ & $1000^3$ & $6.74\times 10^9$ & 22.3  & 5.70 & PlanckDCDM12 & c\\
		PlanckDCDM24\_DMO & 1.0 & $1800^3$ & $1000^3$ & $6.73\times 10^9$ & 22.3  & 5.70 & PlanckDCDM24 & c\\
		\hline
	\end{tabular}
\end{table*}

\begin{table*}
\footnotesize
	\centering
	\begin{threeparttable}[b]
	\caption{The values of the cosmological parameters used in different simulations. Table replicated from \citet{Schaye2023} and expanded with newer models. The columns list the prefix used to indicate the cosmology in the simulation name; the 
 dimensionless Hubble constant, $h$; the total matter density parameter, $\Omega_\text{m}$; the dark energy density parameter, $\Omega_\Lambda$; the
 baryonic matter density parameter, $\Omega_\text{b}$; the sum of the particle masses of the neutrino species, $\sum m_\nu c^2$; the amplitude of
 the primordial matter power spectrum, $A_\text{s}$; the power-law index of the primordial matter power spectrum, $n_\text{s}$; the amplitude of the
 initial power spectrum parametrized as the r.m.s. mass density fluctuation in spheres of radius $8~h^{-1}\,\Mpc$ extrapolated to $z=0$ using linear
 theory, $\sigma_8$; the amplitude of the initial power spectrum parametrized as $S_8\equiv \sigma_8\sqrt{\Omega_\text{m}/0.3}$; the neutrino matter density 
 parameter, $\Omega_\nu \cong \sum m_\nu c^2/(93.14~h^2\,\eV)$, 
and the dark matter decay rate $\Gamma$. Note that the values of the Hubble and density parameters are given at $z=0$. }
	\label{tab:cosmologies}
	\begin{tabular}{lccccccccccc} 
		\hline
		Prefix & $h$ & $\Omega_\text{m}$ & $\Omega_\Lambda$ & $\Omega_\text{b}$ & $\sum m_\nu c^2$  & $A_\text{s}$ & $n_\text{s}$ & $\sigma_8$ & $S_8$ & $\Omega_\nu$ & 	$\Gamma$ \\
         &  &  &  &  &  (eV) &  &  &  &  &  &	$(\rm{H}_0/h)$\\
		\hline
		D3A (fiducial)  & 0.681 & 0.306 & 0.694 & 0.0486 & 0.06 & $2.099\times 10^{-9}$ & 0.967 & 0.807 & 0.815 & $1.39\times 10^{-3}$  & -\\
		Planck          & 0.673 & 0.316 & 0.684 & 0.0494 & 0.06 & $2.101\times 10^{-9}$ & 0.966 & 0.812 & 0.833 & $1.42\times 10^{-3}$ & - \\
		PlanckNu0p12Var & 0.673 & 0.316 & 0.684 & 0.0496 & 0.12 & $2.113\times 10^{-9}$ & 0.967 & 0.800 & 0.821 & $2.85\times 10^{-3}$ & -\\
		PlanckNu0p24Var & 0.662 & 0.328 & 0.672 & 0.0510 & 0.24 & $2.109\times 10^{-9}$ & 0.968 & 0.772 & 0.807 & $5.87\times 10^{-3}$ & -\\
		PlanckNu0p24Fix & 0.673 & 0.316 & 0.684 & 0.0494 & 0.24 & $2.101\times 10^{-9}$ & 0.966 & 0.769 & 0.789 & $5.69\times 10^{-3}$ & -\\
        PlanckNu0p48Fix & 0.673 & 0.316 & 0.684 & 0.0494 & 0.48 & $2.101\times 10^{-9}$ & 0.966 & 0.709 & 0.728 & $11.4\times 10^{-3}$ & -\\
		LS8             & 0.682 & 0.305 & 0.695 & 0.0473 & 0.06 & $1.836\times 10^{-9}$ & 0.965 & 0.760 & 0.766 & $1.39\times 10^{-3}$ & -\\
        PlanckDCDM12    & 0.673 & 0.274 & 0.726 & 0.0494 & 0.06 & $2.101\times 10^{-9}$ & 0.966 & 0.794 & 0.759 & $1.42\times 10^{-3}$ & 0.12\\
		PlanckDCDM24    & 0.673 & 0.239 & 0.726 & 0.0494 & 0.06 & $2.101\times 10^{-9}$ & 0.966 & 0.777 & 0.694 & $1.42\times 10^{-3}$ & 0.24\\
		\hline
	\end{tabular}
	\end{threeparttable}
\end{table*}

%% file: data_products.tex
\section{Data products} \label{sec:data_products}

The FLAMINGO data release comprises a set of data products that are provided consistently across the simulations in the suite, subject to resolution and model-specific differences. A complete description of the file formats, directory layout, dataset definitions, units, and additional metadata is provided in the online documentation\footnote{\url{https://flamingo.strw.leidenuniv.nl/data.html}}. In this section we give a high-level overview of the main data products. 
Table \ref{tab:data_volume} details the storage requirements for each data product per simulation, including the cumulative volume for each category.

\begin{table*}
\footnotesize
\centering
\begin{threeparttable}
    \caption{Approximate sizes of the various data products for different simulations. All values are given in terabytes. The columns list the simulation type (defined by box size, resolution, and whether it is DMO), the size of the $z=0$ snapshot, the size of the $z=0$ SOAP catalogue, the size of a single downsampled HEALPix map (i.e. a single shell or an integrated map), the size of a single halo lightcone, and the size of a single particle lightcone. The final row gives the total data volume of each type of data product over all simulations and all redshifts. The total value for snapshot includes the reduced and downsampled snapshots, the total value for the catalogues includes the HBT-Herons merger trees, and the total values for maps includes the full resolution ($N_{\mathrm{side}} = 16384$) maps.}
    \label{tab:data_volume}
    \begin{tabular}{lD{.}{.}{2}D{.}{.}{3}D{.}{.}{3}D{.}{.}{1}D{.}{.}{1}}
    \hline
    Simulation         & \text{Snapshot} & \text{Halo catalogue} & N_{\mathrm{side}} = 4096~\text{map} & \text{Halo lightcone} & \text{Particle lightcone} \\
    \hline
    L1\_m8             & 4.8      & 0.15           & 0.015                           & 6              & 38                 \\
    L1\_m9             & 0.6      & 0.03           & 0.015                           & 0.6            & 4.4                \\
    L1\_m10            & 0.08     & 0.005          & 0.015                           & -              & 0.5                \\
    L2p8\_m9           & 14       & 0.7            & 0.015                           & 0.6            & 27                 \\
    \hline
    L1\_m8\_DMO        & 1        & 0.05           & 0.002                           & -              & -                  \\
    L1\_m9\_DMO        & 0.1      & 0.008          & 0.001                           & -              & -                  \\
    L1\_m10\_DMO       & 0.02     & 0.001          & 0.001                           & -              & -                  \\
    FLAMINGO-10k       & 21       & 1.8            & -                               & -              & -                  \\
    L2p8\_m9\_DMO      & 2.8      & 0.17           & 0.001                           & -              & -                  \\
    L5p6\_m10\_DMO     & 2.8      & 0.21           & 0.001                           & -              & -                  \\
    L11p2\_m11\_DMO    & 2.8      & 0.19           & -                               & -              & -                  \\
    \hline
    Total              & 1400     & 180            & 600                             & 40             & 160                \\
    \hline
    \end{tabular}

    \end{threeparttable}
\end{table*}

\subsection{Snapshots}

The simulations originally produced 78 snapshots (79 for L1\_m8 and L2p8\_m9, 144 for \flamingo-10k). With the exception of L1\_m9 and L1\_m9\_DMO, we could not retain all snapshots due to data storage constraints. 
Therefore, for each simulation we release the full particle snapshots at the following 
redshifts\footnote{32 snapshots have been kept for \flamingo-10k}:
\begin{equation*}
    z = 5,4,3,2,1.5,1,0.75,0.5,0.4,0.3,0.2,0.1,0.
\end{equation*}

Snapshots are stored in HDF5 format and follow the standard layout of the \swift simulation code. They contain all evolved particle species (dark matter, gas, stars, black holes, and neutrinos), together with extensive metadata describing the cosmology, unit system, and run configuration. Detailed definitions of all fields and their units are given in the online documentation, though this information is also available within the snapshot files themselves. An example of reading snapshot data can be found in \ref{examples_swiftsimio}.

Alongside the full particle snapshots, there are two sets of partial snapshots, which are available for all outputs of the original simulation:
\begin{itemize}
    \item \textbf{Reduced snapshots} contain all the particles within $R_\text{100c}$
    of the centres of a selected sample of massive haloes. The haloes are selected by binning in $M_\mathrm{200c}$, with bins of width 0.05 dex starting at $M_\mathrm{200c}=10^{13} \mathrm{M_\odot}$. If a mass bin contains fewer than 200 haloes, all haloes in that bin are included, otherwise 200 haloes are randomly selected.
    \item \textbf{Downsampled snapshots}\footnote{Downsampled snapshots are not available for the L1m8 simulation.} contain a random 1\% of all particles in the snapshot. The particles retained are selected independently for each snapshot and are therefore not consistent across snapshots. All black hole particles are retained. 
\end{itemize}

Interactive visualizations of slices through \flamingo simulation snapshots are available on the project web site at \url{https://flamingo.strw.leidenuniv.nl/map_slider.html}.

\subsection{Halo and galaxy catalogues}

The history-based structure finder \hbt \citep{Han2018HBT+,Forouhar2025Herons} was used to identify subhaloes. Halo and galaxy catalogues were computed using \soap \citep{mcgibbon25}, which uses the subhalo centres and gravitational boundness of member particles determined by \hbt. The \soap outputs have a file structure similar to that of the snapshots, and can also be read using \swiftsimio. Catalogues are available for each of the original 78 (or 79) simulation snapshots.

\soap provides halo properties defined according to multiple common conventions, allowing users to select the definition most appropriate for their scientific application:
\begin{itemize}
    \item \textbf{Spherical overdensity properties}: The halo radius is defined as the radius at which the mean enclosed density reaches a specified threshold (e.g. $R_\text{200c}$). All particles within this radius are used to compute the corresponding halo properties.
    
    \item \textbf{Inclusive and exclusive spherical apertures}: Properties are measured within apertures of a fixed physical radius. Inclusive apertures include all particles within the aperture, while exclusive apertures include only particles that are gravitationally bound to the subhalo.
    
    \item \textbf{Projected apertures}: Properties are computed within two-dimensional projected physical apertures. For each subhalo, projections are provided along the simulation's $x$, $y$, and $z$ axes.
\end{itemize}

For complete merger tree information, the \hbt catalogues are also provided. The complete documentation can be found on the \hbt website\footnote{\url{https://hbt-herons.strw.leidenuniv.nl/}}.

To facilitate matching haloes across simulations, we construct matching catalogues as follows. For each halo in a reference simulation, we consider all particles bound to the halo. Using their unique particle IDs, we then identify the halo in the comparison simulation that contains the largest number of these particles. This matching procedure is then applied in reverse, allowing bijective matches to be identified. Each hydrodynamical simulation is matched to its corresponding DMO counterpart, which in turn enables matching between different feedback variations.

\subsection{Lightcones}

To facilitate direct comparison with large-area surveys, we release three types of lightcone data products constructed from the simulations: all-sky HEALPix maps, the properties of the particles that contribute to them (``particle lightcones''), and the properties of the haloes contained in them (``halo lightcones'').

Particle lightcones record every particle that crossed the observer’s past lightcone to some maximum redshift. At high redshift the volume and number of particles in the lightcone become extremely large, so for each particle type there is an upper redshift limit beyond which particles are not output. The construction methodology, redshift sampling, and file formats are described in detail in Appendix~A of \citet{Schaye2023} and in the online documentation. 

For each 1~cGpc hydrodynamical simulation, two lightcones corresponding to different observer positions are available, with HEALPix maps for one extending to $z=3$ and for the other to $z=0.5$ (see Table~A1 of \citealt{Schaye2023} for the extent of the various particle lightcones). The L2p8 simulation originally produced eight lightcone outputs, so we provide eight sets of maps, although only two particle lightcones remain. For this simulation, each lightcone's HEALPix map extends to $z=5$. See the online documentation for the redshift ranges covered by the particle lightcones for the different simulations and particle species. 

We release halo lightcones that extend to redshift $z=15$. These are constructed using black hole particles as tracers of the subhalo positions, with halo properties obtained from the \soap catalogues at the nearest snapshot in redshift.

For the DMO simulations there are HEALPix maps but no particle or halo lightcones.

\subsubsection*{HEALPix maps}

We release full-sky projected maps in concentric spherical shells around the observer. When a particle crosses the observer’s lightcone, we determine in which radial shell it lies and then accumulate its contribution to the appropriate maps. These maps use the HEALPix pixelisation scheme \citep{Gorski2005} to divide the sky into pixels of equal area. The maps are provided at two resolutions. The highest resolution maps have an angular resolution of $\approx 13$ arcseconds ($N_{\mathrm{side}} = 16384$). As these maps can be computationally demanding to work with, we also provide downsampled maps with an angular resolution of $\approx 50$ arcseconds ($N_{\mathrm{side}} = 4096$). 

The widths of the shells used to construct the FLAMINGO HEALPix maps can exceed the size of the simulation box. This means that a photon travelling towards the observer may pass through multiple periodic replications of the same structure, introducing spurious correlations in the maps.
We therefore also release a number of integrated maps for which the FLAMINGO shells have been randomly rotated whenever the lightcone diameter exceeds a multiple of the box size. The same rotation has been applied to each of the observables, meaning the different integrated maps can be directly compared and cross correlated.

The following HEALPix maps are currently available:

\begin{itemize}
    \item \textbf{Mass} \citep[][Appendix~A2.3]{Schaye2023}. The total projected mass in each pixel for different particle types. There are individual maps for black hole mass, dark matter mass, stellar mass, gas mass (both with and without smoothing), neutrino mass, and total mass. \textbf{[Shells only]}
    
    \item \textbf{Star formation rate} \citep[][Appendix~A2.3]{Schaye2023}. The total projected star formation rate in each pixel. Each time a gas particle crosses the lightcone its associated star formation rate is accumulated to the pixel containing the particle’s position on the sky. \textbf{[Shells only]}
    
    \item \textbf{Weak lensing convergence maps} \citep{broxterman24}. Constructed from the total matter lightcone shells using a backward ray-tracing methodology and assuming a simple analytic non-tomographic \textit{Euclid}-like source redshift distribution ($N_\mathrm{side}$ = 8192). \textbf{[Integrated only]}
    
    \item \textbf{CMB lensing} \citep{Yang2025}. Lensing convergence maps using the matter overdensity computed from the total-matter lightcone outputs. The two-dimensional projected overdensity map shells are integrated along the line of sight up to $z = 3.0$ and 5.0 for the 1 and 2.8~cGpc simulation volumes, respectively, weighted by the CMB lensing kernel. \textbf{[Integrated only]}
    
    \item \textbf{Thermal Sunyaev-Zel’dovich effect} \citep{Schaye2023,Yang2025}. Constructed by accumulating the Compton-$y$ parameter. In addition to the maps corresponding to individual redshift shells, integrated maps are available, where lensing effects are included using \textsc {pixell}\footnote{\url{https://github.com/simonsobs/pixell}}, applied shell by shell with the integrated convergence map up to the shell of interest. Additionally, a relativistically corrected version of the thermal SZ intensity fluctuation map is available, with the correction function computed using \textsc{SZpack} \citep{SZpack_ref1,Szpack_ref2}. Contributions from gas particles that are star forming or have recently been directly heated\footnote{A gas particle is considered recently heated if it has been directly heated by AGN feedback within the last $15~\mathrm{Myr}$ and has a temperature of $10^{-1} \Delta T_{\mathrm{AGN}} < T < 10^{0.3} \Delta T_{\mathrm{AGN}}$, where, for the fiducial model at m9 resolution, $\Delta T_{\mathrm{AGN}}=10^{7.78}\mathrm{K}$.} by AGN feedback are excluded. \textbf{[Integrated and shells]}
    
    \item \textbf{Kinetic Sunyaev-Zel’dovich effect} \citep{Schaye2023,Yang2025}. Constructed by accumulating the Doppler-$b$ parameter. Integrated maps are available, where the mapping between the Doppler parameter is given by $\Delta T_{\rm kSZ}/T_{\rm CMB} = -b$, where $T_{\rm CMB}=2.73~$K. Lensing effects are computed using the same procedure as used for lensing of the thermal SZ map. As for the thermal SZ effect, contributions from particles that are star forming or have recently been heated by AGN are excluded. \textbf{[Integrated and shells]}
    
    \item \textbf{Anisotropic screening (optical depth $\tau$)} \citep{Yang2025}. The screening effect is proportional to the line-of-sight integrated optical depth $\tau(\mathbf{\hat{n}})$, which is given by the Thomson scattering cross-section multiplied by the stacked dispersion measure map. As for the thermal SZ effect, contributions from particles that are star forming or have recently been heated by AGN are excluded.  \textbf{[Integrated and shells]}
    
    \item \textbf{Dispersion measure} \citep[][Appendix~A2.3]{Schaye2023}. Computed by integrating the free electron number densities in the lightcone outputs. As for the thermal SZ effect, contributions from particles that are star forming or have recently been heated by AGN are excluded. \textbf{[Shells only]}
    
    \item \textbf{Cosmic infrared background} \citep{Yang2025}. Generated from the star formation rate lightcone outputs, with the bolometric infrared luminosity assumed to be proportional to the star formation rate \citep{LIR_SFR_relation}. The luminosity at a given frequency is then computed using a greybody radiation spectral energy distribution (SED) for infrared sources \citep{P16_CIB_SED_template}, with the SED parameters determined by fitting to the measured 353, 545, and 857 GHz CIB power spectra from \citet{L19_CIB}. The same procedure as used for lensing of the thermal SZ map is applied here. \textbf{[Integrated only]}
    
    \item \textbf{Radio point source emission} \citep{Yang2025}. Constructed from the black hole particle lightcone outputs. Radio luminosities are assigned by abundance matching the bolometric AGN luminosity function to the LOFAR 150 MHz luminosity function up to $z=2.5$ \citep{LOFAR_RLF}. The lensed source fluxes are extrapolated to higher CMB frequencies using a power-law SED, with the power index fitted to match the measured radio source counts from the SPT survey at 95, 150, and 220~GHz \citep{Everett_SPT}. The procedure is repeated for subsamples selected by black hole accretion state using different Eddington ratio cuts, $\lambda_{\rm Edd} \equiv L_{\rm bol}/L_{\rm Edd}$: $\lambda_{\rm Edd}<10^{-2}, 10^{-3} ~\textrm{and}~10^{-6}$. \textbf{[Integrated only]}
    
    \item \textbf{Diffuse X-ray emission} \citep{Schaye2023,McDonald2026}. Constructed by summing the X-ray fluxes from gas particles with temperature $10^{5} \le T/\text{K} \le 10^{9.5}$ in three different energy bands: eROSITA high ($2.3-8.0$~keV), eROSITA low ($0.2-2.3$~keV) and ROSAT ($0.5-2.0$~keV). In each band, the X-ray fluxes in each pixel are available in units of $\text{photons}\,\text{cm}^{-2}\,\text{s}^{-1}$ and $\text{erg}\,\text{cm}^{-2}\,\text{s}^{-1}$,
    or $\text{photons}\,\text{s}^{-1}$ and $\text{erg}\,\text{s}^{-1}$ when convolved with the response matrix of the telescope's energy bands.
    We utilise the effective area information from the publicly available ROSAT\footnote{\url{https://heasarc.gsfc.nasa.gov/docs/rosat/ruh/handbook/node122.html\#tabeffarea}} on-axis response function and the survey-averaged eROSITA\footnote{Specifically TM8, which excludes TM5 and TM7 as they suffer from light leakage \citep[see][]{Predehl_2021}, \url{https://erosita.mpe.mpg.de/dr1/eSASS4DR1/eSASS4DR1_arfrmf/}} auxiliary response files. Each particle's X-ray emissivity is computed using tables generated with CLOUDY \citep[][version 17.02]{Ferland2017}, accounting for the individual elemental abundances \citep[see][for details]{Braspenning2024} and is consistent with the radiative cooling rates used during the simulation \citep{Ploeckinger2020}. As for the thermal SZ effect and dispersion measure, X-ray emission from gas particles that are star forming or that have recently been heated by AGN feedback is excluded. \textbf{[Integrated and shells]}
    
    \item \textbf{X-ray emission from AGN} \citep{McDonald2026}. X-ray flux from AGN as predicted directly by \flamingo's black hole accretion rates or using the fiducial AGN abundance matching model of \citet{McDonald2026} for the same bands as the diffuse X-ray emission. These maps will become available upon acceptance of \citet{McDonald2026}. \textbf{[Integrated only]}
\end{itemize}

Interactive visualizations of the HEALPix maps are available on the \flamingo project web site, at \url{https://flamingo.strw.leidenuniv.nl/lightcone_slider_CY.html}.

\subsection{Power spectra}

We release various 3-D power spectra computed on-the-fly at 123 redshifts. For each redshift, we provide the comoving wavenumber, $k$, and the corresponding shot-noise subtracted power, $P(k)$, for 14 distinct auto- and cross-spectra. These include power spectra for total matter, cold dark matter (CDM), gas, the sum of stars and black holes (starBH), and electron pressure. We also provide cross-power spectra for CDM with gas, neutrinos, and starBH; gas with total matter, neutrinos, and starBH; total matter with electron pressure; starBH with neutrinos; and the cross-correlation between two randomly selected subsets of the neutrino particles, which is done for shot-noise suppression. Note that the baryon response of the power spectrum (i.e., the ratio of the total matter power spectrum in the hydrodynamical simulation to its dark-matter-only counterpart) was already released, including in the form of a Gaussian process emulator embedded in a \textsc{python} package, by \cite{Schaller2025}.

\subsection{Initial conditions}

The initial conditions for the L1\_m9 and L1\_m9\_DMO simulations are available on the data release website. These are using the \swift format for initial conditions based on the \texttt{hdf5} storage library. For all the other simulations in the \flamingo suite, we intend to provide parameter files for the \textsc{monofonIC} code that can be used to reproduce the initial conditions. Further details on the creation of the initial conditions can be found in Section~2.4 and Appendix~B of \citet{Schaye2023}.

Both \textsc{monofonIC} and the \swift simulation code (including all physics modules used in the \flamingo simulations) are open source. We have also published the parameter files used to run the simulations. This means that, in principle, anyone will be able to reproduce the simulations using the public code and parameters. Users could run their own \flamingo variants in different volumes or with different model parameters, for example, or run zoom simulations of individual objects.

\subsection{Future data products}

We anticipate that additional derived and value-added products will be computed and released over time as the analysis of the simulation continues. New data products will follow the same directory and format conventions as the existing release where possible, and will be documented online. Announcements of new releases and updates will be made via a dedicated mailing list. Instructions for subscribing can be found at \url{https://flamingo.strw.leidenuniv.nl/mailing-list.html}.

%% file: data_access.tex
\section{Data access} \label{sec:data_access}

In this section we describe the system we have set up to provide access to the \flamingo simulations.

\subsection{Design considerations}

The \flamingo simulations consist of a large and varied set of outputs. The total size of the data release is in excess of 2.3~PB and includes simulation snapshots, halo catalogues, all-sky HEALPix lightcone maps, and lightcone particle data. The complete data set and even individual simulation snapshots are too large for most users to download in full. We therefore provide a mechanism that allows for selective access to the data to enable scientific exploitation of the simulations by users who might have limited access to computing resources.

The \flamingo outputs are stored on the DiRAC Memory Intensive system in Durham, where they are accessed by researchers running HPC jobs. The large size of the outputs means that duplicating them (for example, to cloud storage) is not desirable. This implies that the data must be served directly from the Lustre parallel file system where they are stored, and they must remain in the HDF5 file format required for analysis by local HPC users. This also precludes the use of a relational database, as in the Millennium \citep{2005Natur.435..629S} and EAGLE \citep{Schaye2015} projects, which would require conversion of the data into the database server's internal representation.

The HDF5 file format was originally designed for local disk storage and is not well suited to remote access. In particular, metadata can be distributed throughout the file in many small chunks. This means that reading HDF5 files using generic tools for remote file system access can be inefficient because it requires many small, random access reads. This problem can be addressed by rewriting the files as "cloud optimized" HDF5 or NetCDF, where HDF5 library parameters are tuned to consolidate metadata as far as possible, or using a format specifically optimized for remote access, such as Zarr\footnote{\url{https://github.com/zarr-developers/}}. However, for the \flamingo data release we wish to serve our existing HDF5 data. We also have some idea of likely access patterns: cutting out regions from the snapshots or patches on the sky from lightcone outputs will involve requesting many slices from one or two dimensional datasets. This can be efficiently implemented as a service which carries out the necessary random access I/O operations on the local file system and produces a compact response for the user to download.

For users who do have enough network bandwidth and storage to download complete outputs, we wish to support full file downloads. The system hosting the data has a Globus\footnote{\url{https://globus.org}} endpoint which can be used for bulk data transfers, although this currently requires authentication with a local user account. For simplicity, we would like to also support plain http downloads through the same server that provides selective access to the data.

Additional considerations include:
\begin{itemize}
    \item From experience with previous simulation data releases, we know that this type of service typically has small numbers of concurrent users. Multiple simultaneous requests must be supported, but scaling beyond a single server is not required.
    \item The datasets of most interest to users are generally large numeric arrays. Metadata must also be accessible, but fast access to large arrays of fixed size data types is a priority.
    \item We do not expect the simulation data to change after release, although new data products may be added later.
\end{itemize}

\subsection{Implementation}

Given the above constraints, we chose to set up a relatively simple, lightweight service capable of serving the contents of HDF5 files stored on a local file system. Users send http requests specifying a file, a dataset of interest, or possibly a slice within the dataset, and the server reads and returns the corresponding data. This service is designed to re-use existing DiRAC infrastructure as far as possible.

\subsubsection{Serialization format}

Returning data in HDF5 format poses some practical problems. If we wish to construct and return a new HDF5 file with the requested data, it is difficult to avoid caching the full response on the server before returning any data to the user. This would delay the response and require potentially large and unpredictable amounts of temporary storage. This isn't an insurmountable problem (the IllustrisTNG web service is able to construct HDF5 responses, for example) but we prefer to choose a format better suited to streaming so that the user can begin downloading data while it is still being read from the underlying storage. We anticipate that most requests will be for large arrays of numeric data, so the response format must be able to represent these arrays efficiently.

We opted to encode responses using the messagepack\footnote{\url{https://www.msgpack.org}} serialization format for the following reasons: Messagepack is a compact, binary format that provides all the data types necessary to represent the metadata in our HDF5 simulation outputs. It also includes a raw binary buffer data type which can be used to represent the body of large arrays in an efficient way. This means that dataset values read from HDF5 files can be copied directly to the response stream without translation. Additionally, the messagepack format is much simpler to implement than HDF5 and so many implementations exist.

Note that we do not intend for users to interact with messagepack responses directly. We provide a python module which can unpack responses into \texttt{numpy} arrays, which can be either used directly in calculations in memory or written to local HDF5 files using \texttt{h5py}.

\subsubsection{Parallelization}

The \flamingo simulation outputs make extensive use of HDF5 filters for data compression. This means that we cannot easily implement our own routines to read data from the files, we must use the HDF5 library. This is a problem because the library does not support thread-based parallelism. Multiple threads may call the library, but these calls are serialized using a single, global lock. Since our service needs to handle multiple concurrent requests, it is necessary to handle each request in a separate process.

We implemented a process pool to handle requests in parallel. When the server receives a request, a process from the pool is assigned to read the data. The process uses the HDF5 library to read the requested data in chunks up to some maximum buffer size. Each chunk is serialized and passed back to the server via shared memory. The server can then write each chunk to the response stream.

To reduce load on the file system, the processes cache recently opened files and datasets, and the server routes requests to processes that already have the requested file and dataset in their cache where possible.

\subsubsection{Server implementation}

The server is a single instance of the Apache Tomcat\footnote{\url{https://tomcat.apache.org/}} application server running on a system with 24 CPU cores, 700~GB memory and a 20~Gbit/s network connection. While this system will not scale to very large numbers of users, it is capable of providing high throughput for individual requests. Large HDF5 datasets can be streamed at up to 200~MB/s, typically limited by the rate at which a single process can read from the file system, and likely faster than most users can download the response.

The server responds to two types of requests for HDF5 data:
\begin{itemize}
    \item Recursive serialization of an HDF5 group and its contents, omitting dataset contents larger than some size threshold and subject to a maximum depth for recursion into sub-groups. 
    \item Serialization of all or part of the contents of a dataset.
\end{itemize}
The first request type can be used to determine what datasets are available, retrieve metadata and efficiently fetch any small datasets. The second type is used to retrieve the actual simulation data. The server also supports requests for file and directory metadata and we have reused code developed for the release of the EAGLE simulations to allow full downloads of files and directories.

The source code for our server implementation is available on github\footnote{\url{https://github.com/jchelly/hdfstream-api}}.

\subsubsection{Python client module}

We implemented a python client module\footnote{\url{https://hdfstream-python.readthedocs.io/}} that can be used to request data from the server. This is designed to resemble the popular \texttt{h5py}\footnote{\url{https://docs.h5py.org}} module for ease of use. HDF5 groups are represented as dict-like containers of datasets and sub-groups. HDF5 datasets are represented by "remote dataset" objects, which are array-like objects that request data from the server when their elements are accessed. These remote datasets support a subset of \texttt{numpy} array indexing syntax. Simple slices, specifying a contiguous range of elements in each dimension, generate a request for the corresponding elements of the dataset on the server and return a \texttt{numpy} ndarray. A remote dataset may also be indexed with an integer or boolean array in the first dimension. In this case, the index array is translated into a sequence of contiguous slices which are retrieved from the server with a single http request. In \ref{examples_soap} we illustrate how this module can be used to access \flamingo data.

We also adapted the \swiftsimio module \citep{Borrow2020} to access \flamingo snapshots and halo catalogues using our web service. \swiftsimio is designed to provide an easy-to-use, high-level interface for working with snapshots generated by the \swift simulation code \citep{SWIFT}. The example in \ref{examples_swiftsimio} shows how to open a snapshot on the server using \swiftsimio and \ref{examples_cutout} shows how to download the particles in some region of the simulation volume and save them to a local snapshot file for later analysis. In \ref{examples_halo_particles} we show how to select a halo from the SOAP catalogue and download the particles bound to it.

We plan to adapt the widely used \codename{pynbody} module \citep{2013ascl.soft05002P} to access \flamingo in a similar way. 

\subsubsection{Web interface}

We created a web interface that allows users to browse the contents of the data release. It presents a virtual directory structure containing the simulation data and can also display the contents of HDF5 files. This latter capability is useful because the \flamingo outputs contain many descriptive attributes. The web interface also presents full download links for the data and extensive documentation. The data browsing facility is implemented as a javascript client, which makes requests to the server for file and directory contents to display using the same data access API as the python module.

The source code for the web interface, including documentation describing the \flamingo outputs, is also published on github\footnote{\url{https://github.com/jchelly/flamingo-docs}}.

%% file: discussion.tex
\section{Discussion} \label{sec:discussion}

\subsection{(Dis)agreement with observations and scientific caveats}

The \flamingo universe is not the real Universe. Nevertheless, the fiducial \flamingo model has been shown to reproduce many observations. On the galaxy population side this includes the $z=0$ galaxy stellar mass function, black hole mass -- stellar mass relation, specific star formation rates, and metallicities, though not necessarily over the full mass range covered by the catalogues \citep{Schaye2023}. While all resolutions reproduce the observed $z=0$ galaxy stellar mass function in the mass range for which they were calibrated, they diverge not only below their resolution limits, but also above the upper mass limit used for the calibration, $M_* = 10^{11.5}\,$M$_\odot$, where it also becomes sensitive to the chosen aperture \citep{Schaye2023}. 

Only a few studies have so far compared the predicted galaxy populations to observations at higher $z$. The cosmic star formation history is reproduced up to the redshifts where unresolved low-mass galaxies begin to dominate ($z \approx 5$ for m9 resolution) \citep{Schaye2023}. \citet{Lim2024,Kumar2025} showed that \flamingo reproduces high-star formation rate objects such as submillimeter galaxies better than previous simulations. 

Discrepancies with galaxy observations include $z=0$ galaxy sizes being slightly too large for $M_* \lesssim 10^{10}\,$M$_\odot$ and quenched fractions being too low for the most massive galaxies, $M_* \gtrsim 10^{12}\,$M$_\odot$ \citep{Schaye2023}. At high $z$, \citet{Baker2025} found that, similar to most other simulations, \flamingo predicts too few massive, quiescent galaxies. 

The quasar luminosity function compares well with observation for $z \lesssim 1$, but its bright end is underestimated at higher redshift \citep{Ding2025}. The clustering of quasars is reproduced for $z \lesssim 3$, but underestimated at $z\approx 4$  \citep{Ding2025}. 

Galaxy cluster scaling relations, including their evolution, agree well with the data \citep[e.g.][]{Schaye2023,Braspenning2024}, even for masses greater than the (box-size dependent) upper halo mass limit used in the calibration ($M_\text{500c} > 10^{13.73}$ and $10^{14.36}\,$M$_\odot$ for the m8 and m9 resolutions, respectively). Cluster thermodynamic radial profiles are also reasonable, also at high redshift, though metallicities (i.e.\ iron abundances) are substantially too high at small radii \citep{Braspenning2024}. 

In the group/low-mass cluster  regime ($10^{13} <  M_\text{500c}/\text{M}_\odot < 10^{14}$) the situation is currently unclear. Stacks of kinetic SZ observations combined with weak lensing mass measurements, and stacks of X-ray data from optically-selected groups are both fit much better by the low gas fraction variations of \flamingo than the fiducial model \citep[e.g.][]{McCarthy2025,Hadzhiyska2025,Popesso2025,Kovac2025,Siegel2025,Bigwood2025b}, suggesting that feedback in the regime of groups or their progenitors needs to be rather strong. However, studies using X-ray observations of individually detected groups, some of which account for selection effects, indicate that the fiducial \flamingo model is consistent with the data, but the strong feedback variants are not \citep{Kugel2023,Eckert2025}. This apparent contradiction could indicate systematic errors or selection effects in some of the data sets, but it could also indicate that the simulations are insufficiently realistic because kinetic SZ and X-ray observations probe different redshifts and radii. 

On the large-scale structure side, CMB lensing and its cross correlations with the thermal SZ effect and cosmic shear agree with the simulations \citep{McCarthy2023,Schaye2023}, while, for the fiducial cosmology, the cosmic shear and thermal SZ power spectra show well-known tensions \citep[reflecting the so-called $S_8$ tension;][]{McCarthy2023}. However, recent observations of cosmic shear \citep{Wright2025} and the thermal SZ power spectrum \citep{Efstathiou2025} are discrepant with earlier ones and, at least on large scales, are closer to the \flamingo predictions. The cross correlation between X-ray emission and cosmic shear is consistent with the data \citep{McDonald2026}. The cosmic infrared background and its cross correlation with CMB lensing also reproduce the data \citep{Yang2025}. 

\subsection{On tests of numerical convergence}
The \flamingo simulations include three different numerical resolutions, which differ by factors of 8 (2) in mass (spatial) resolution. For each resolution the subgrid feedback was independently calibrated to match the same observables, though over different galaxy and cluster mass ranges \citep{Kugel2023}.

The resolution limit will generally depend on the property that is being measured and the level of precision required. We advise testing convergence with numerical resolution for each question that is investigated. However, we stress that differences between different resolutions may not (only) reflect direct resolution effects. This is because the calibration is imperfect and, more importantly, because the calibration is performed only for a few observables. Simulations using different resolutions are effectively different physical models because they resolve different physical processes and because important aspects of the feedback, such as the mass heated/kicked per event, depend directly on the resolution. Because the physical models differ, different resolutions cannot be expected to necessarily produce similar results for observables that were not considered during the calibration. For example, in the left panel of Fig.~8 of \cite{Schaye2023} we can see that the galaxy stellar mass function at a mass of around $10^{12}$M$_\odot$ is significantly different between m8, m9 and m10 resolutions. Similarly, the stellar half mass radii of active galaxies, shown in Fig.~14 of the same paper, depend on the resolution of the simulation.

\subsection{Bugs and issues affecting convergence tests}

The interpretation of resolution tests for \flamingo is complicated by the fact that we incorporated one improvement and fixed one bug in those simulations that were run in a later phase of the project.

In all intermediate-resolution simulations except for the Jet models, particles with metallicity equal to precisely zero used a star formation threshold density of $n_\text{H} = 10~\text{cm}^{-3}$ instead of  $10^{-1}~\text{cm}^{-3}$. Tests show that this only has significant effects on galaxies with fewer than 10 star particles, where it artificially suppresses the stellar-to-halo mass ratio.

A more important difference between the non-Jet m9 simulations and all other models is related to the black hole repositioning scheme. Because the simulations do not resolve the dynamical friction experienced by black holes, they are moved down the gradient of the gravitational potential by hand. For the purpose of this repositioning of black hole particles, we should\footnote{As far as we know, this subtraction was not done by previous cosmological simulations.} subtract the contribution from the black hole to the gravitational potential. As already reported in \citet{Schaye2023}, this was, however, only done for the m8 and m10 resolutions and for the Jet variations of the set using m9 resolution. In all other simulations the black hole particles are repositioned less efficiently because they can become trapped in their own potential wells. As shown in \citet{Bahe2022}, less efficient repositioning reduces the efficiency of AGN feedback. Tests in smaller volumes show that this is the primary cause of the higher quenched fractions among massive galaxies in the fiducial m8 and the Jet m9 simulations compared to the other m9 models.

\subsection{Other bugs and issues}
The following significant bugs and issues were already discussed in \citet{Schaye2023}, 
\begin{itemize}
\item For computational efficiency reasons, black hole particles are only repositioned (i.e.\ moved by hand down the potential gradient to compensate for unresolved dynamical friction) onto gas particles. For gas-poor galaxies, such as low-mass satellites, this can have the consequence that black holes leave their host galaxy, either temporarily or permanently. Care should therefore be taken when studying black holes and/or AGN feedback in satellite galaxies. This issue also affects the galaxy lightcones. Because the locations of black hole particles are used to place galaxies on the lightcone, a fraction of poorly resolved galaxies, in particular satellites, is missing from the galaxy lightcones (see the online documentation for quantitative information). 

\item AGN feedback is implemented by heating/kicking particles to very high temperatures/velocities, which is necessary to overcome numerical overcooling \citep[e.g.][]{DallaVecchia2012}. Because the gas particles subject to energy injection by feedback are selected from the SPH neighbours of black holes/young stars, they tend to be part of the dense interstellar medium. This implies that for a few time steps following energy injection, i.e.\ until the particles have responded hydrodynamically to the energy injection, such dense and hot gas can artificially distort the observational properties of galaxies, such as their X-ray emission. We therefore advise to test the effect of excluding recently heated/kicked particles, which can be done using the particle property tracking the last time a particle was injected with AGN feedback energy. For some observables (gas and spectroscopic-like temperatures, Compton-$y$, X-ray) the SOAP catalogues provide versions that exclude particles that were subject to direct AGN heating in the last 15~Myr and whose temperatures are between $10^{-1}\Delta T_\text{AGN}$ and $10^{0.3}\Delta T_\text{AGN}$, where $\Delta T_\text{AGN}$ is the AGN heating temperature. However, due to a bug, the SOAP catalogues of all simulations assumed the same value of $\Delta T_\text{AGN} = 10^{7.95}\,$K, which is the AGN heating temperature used in the m9 simulations with the fiducial galaxy formation model, instead of the values listed in Table~1 of \citet{Schaye2023}, which vary between $\Delta T_\text{AGN} = 10^{7.71}\,$K (for fgas$+2\sigma$) and $10^{8.40}\,$K (for fgas$-8\sigma$). 

\item The original HEALPix maps for the kinetic SZ effect and the dispersion measure were computed using a wrong power of the cosmological scale factor. This was corrected using the expansion factor of the midpoint of each lightcone shell, which has a width of $\Delta z=0.05$. Note that for those simulations and redshifts for which particle lightcone data is available, the maps can be recomputed if desired (the particle lightcone data has already been corrected). 
\end{itemize}

A comprehensive list of all known issues and bugs, including those mentioned above, as well as additional ones with minor effects, can be found at \url{https://flamingo.strw.leidenuniv.nl/known-issues.html} and will be updated as more issues are discovered.

\section{Acknowledgement of usage} \label{sec:credits}
Publications making use of the public FLAMINGO data are kindly requested to:
\begin{itemize}
    \item Cite the two papers presenting the FLAMINGO project: \citet{Schaye2023} and \citet{Kugel2023}, as well as this data release paper.
    \item When referring to specific aspects of the methodology, cite the original papers presenting those aspects rather than the overview paper.
    \item Add the following statement to the acknowledgements: \textit{We acknowledge the Virgo Consortium for making their simulation data available. The \flamingo simulations were performed using the Durham Memory Intensive system managed by the Institute for Computational Cosmology on behalf of the STFC DiRAC facility (\url{www.dirac.ac.uk}).}

\end{itemize}

%% file: conclusions.tex
\section{Conclusions} \label{sec:conclusions}
The \flamingo suite of cosmological simulations is a set of 22 hydrodynamical and 16 dark-matter-only simulations in volumes $\ge 1~\rm{cGpc}^3$. These simulations were designed as virtual counterparts to late-time Universe cosmological probes, and thus provide a test bed for the analysis pipelines, especially to understand the effects of systematics induced by baryonic physics. Besides cosmological applications, the \flamingo runs provide a theoretical framework for the study of rare objects, such as clusters of galaxies. The fiducial simulations have been carefully calibrated, via machine-learning techniques, to the low-$z$ galaxy stellar mass function and the gas fractions of clusters, offering a plausible realistic baseline scenario. Importantly, the fiducial models are complemented with resolution variations, variations in the baryon calibration, and in the assumed cosmology. 

We release here $> 2.3$ petabytes of data from this simulation suite, which we make accessible via a web interface. In addition to the raw data, we provide catalogues of haloes and galaxies, power spectra of various quantities, full-sky tomographic and integrated lightcone maps, and particles data from which additional lightcone maps can be computed. By making these data publicly available, we hope to provide the community with a valuable resource for the theoretical interpretation of cosmology and galaxy evolution datasets.

%% file: examples.tex
\section{Usage examples}
\label{examples}

\subsection{Remote access to HDF5 files}
\label{examples_soap}

In this section we illustrate how to use the \flamingo web service to access the contents of HDF5 files on the server. The python module used to access the service can be installed using pip:

\begin{lstlisting}[style=code]
pip install hdfstream
\end{lstlisting}

We can connect to the service in a python session as follows:

\begin{lstlisting}[style=code, language=Python]
import hdfstream
root_dir = hdfstream.open("cosma", "/")
\end{lstlisting}

Here, \verb|"cosma"| is an alias for the server to connect to and we are opening the root directory on the server. The result is a \verb|RemoteDirectory| object which represents the service root directory. Files and sub-directories are opened by subscripting the \verb|RemoteDirectory| with a path:

\begin{lstlisting}[style=code, language=Python]
flamingo_dir = root_dir["FLAMINGO"]
\end{lstlisting}

This returns a \verb|RemoteFile| or \verb|RemoteDirectory| representing the file or directory at the specified path. Since a \verb|RemoteDirectory| behaves like a dictionary of files and subdirectories, we can obtain a directory listing by printing its keys:

\begin{lstlisting}[style=code, language=Python]
print(list(flamingo_dir))
\end{lstlisting}

If we already have the location of a file of interest (from the web interface, for example), we can open it by subscripting the root directory object with the full path. For example, a SOAP halo catalogue can be opened with:

\begin{lstlisting}[style=code, language=Python]
soap_file = root_dir["FLAMINGO/L1_m10/L1_m10_DMO/SOAP-HBT/halo_properties_0077.hdf5"]
\end{lstlisting}

This opens the file containing the halo catalogue for the redshift $z=0$ snapshot of the \verb|L1_m10| simulation. It returns a \verb|RemoteFile| object, which is a dict-like container for HDF5 groups and datasets. It behaves similarly to an \texttt{h5py} file object, but dataset elements are downloaded on access. We can download the positions of all haloes in the snapshot with:

\begin{lstlisting}[style=code, language=Python]
halo_pos = soap_file["BoundSubhalo/CentreOfMass"][...]
\end{lstlisting}

Here, indexing the dataset with an ellipsis triggers a download of the full dataset contents and the result is a \texttt{numpy} array. We can also download parts of datasets:

\begin{lstlisting}[style=code, language=Python]
halo_pos = soap_file["BoundSubhalo/CentreOfMass"][0:1000,:]
\end{lstlisting}

This would download just the first 1000 elements in the first dimension. The units of the dataset can be determined by inspecting its attributes:

\begin{lstlisting}[style=code, language=Python]
print(soap_file["BoundSubhalo/CentreOfMass"].attrs)
\end{lstlisting}

This will return details which include a description of the dataset, a conversion factor to CGS units, and any dependence on the cosmological expansion factor, in the case of co-moving quantities. This approach can be used to access all types of \flamingo data products. However, we have also adapted the \swiftsimio module to provide a higher level interface to snapshots and SOAP halo catalogues.

\subsection{Accessing snapshots with \swiftsimio}
\label{examples_swiftsimio}

The \swiftsimio python module can be used to access datasets in \flamingo snapshots and halo catalogues, either by reading downloaded files directly with \texttt{h5py} or by accessing files on the server using the \hdfstream module. It handles metadata by defining a \verb|numpy.ndarray| sub-class which contains information about the units of each dataset, whether the quantity is stored in co-moving coordinates, and any dependence on the cosmological scale factor. It can be installed with:

\begin{lstlisting}[style=code]
pip install swiftsimio hdfstream
\end{lstlisting}

The snapshots and halo catalogues are stored in very similar formats and \swiftsimio is capable of reading both. Here, we use the redshift zero snapshot from the \verb|L1_m10| simulation as an example. The file containing the snapshot can be opened in python with:

\begin{lstlisting}[style=code, language=Python]
import hdfstream
root_dir = hdfstream.open("cosma", "/")
snap_file = root_dir["FLAMINGO/L1_m10/L1_m10/snapshots/flamingo_0077/flamingo_0077.hdf5"]
\end{lstlisting}

We can then use \swiftsimio to access the contents of the file:

\begin{lstlisting}[style=code, language=Python]
import swiftsimio as sw
snapshot = sw.load(snap_file)
\end{lstlisting}

This returns a \verb|SWIFTDataset| object, which has attributes corresponding to the particle types in the snapshot:
\begin{lstlisting}[style=code, language=Python]
snapshot.gas
snapshot.dark_matter
snapshot.stars
snapshot.black_holes
snapshot.neutrinos
\end{lstlisting}
Each of these particle types has attributes corresponding to particle properties and accessing an attribute triggers a request to the server for the data. For example, we can download the positions of the star particles with:
\begin{lstlisting}[style=code, language=Python]
star_pos = snapshot.stars.coordinates
\end{lstlisting}
The result is a \swiftsimio cosmo array, which is an array with additional metadata describing the units of the quantity and any dependence on the cosmological scale factor. In this case, it records that coordinates are stored in comoving Mpc:
\begin{lstlisting}[style=code, language=Python]
>>> print(star_pos)
[[649.22656783 524.32466783 595.55844783]
 ...
 [987.07910218 995.73126218  88.39935218]] Mpc (Comoving)
\end{lstlisting}

\subsection{Downloading regions of interest}
\label{examples_cutout}

The particles in \swift snapshots are ordered in a way that makes it possible to read regions from a snapshot without reading the entire file. The simulation volume is divided into grid cells and the particles are sorted according to which cell they fall in. Given a range of coordinates in $x$, $y$, and $z$, \swiftsimio can compute which parts of the file to read or download to find the corresponding particles. The example below shows how to do this:

\begin{lstlisting}[style=code, language=Python]
# Open the root directory
import hdfstream
root_dir = hdfstream.open("cosma", "/")

# Open the snapshot file.
snap_file = root_dir["FLAMINGO/L1_m10/L1_m10/snapshots/flamingo_0077/flamingo_0077.hdf5"]

# Create a mask object
import swiftsimio as sw
mask = sw.mask(snap_file)

# Define the region to read
from unyt import Mpc
load_region = [[100*Mpc, 150*Mpc], [100*Mpc, 150*Mpc], [100*Mpc,150*Mpc]]

# Constrain the region to read
mask.constrain_spatial(load_region)

# Open the snapshot using the mask
snap = sw.load(snap_file, mask=mask)

# Download coordinates of gas particles in the region
gas_pos = snap.gas.coordinates
\end{lstlisting}

It is also possible to save the region as a new snapshot file. Using the \texttt{mask} object from the example above:

\begin{lstlisting}[style=code, language=Python]
sw.subset_writer.write_subset("flamingo_region_0077.hdf5", mask)
\end{lstlisting}

This will download all particles in the selected region of the snapshot and save them to a new HDF5 file on the user's local file system.

\subsection{Downloading particles in a halo}
\label{examples_halo_particles}

The python code below shows how to identify a halo of interest in a SOAP halo catalogue and download the particles which belong to that halo, using the most massive halo in the \verb|L1_m10_DMO| simulation at redshift $z = 0$ as an example. This is done by reading all particles in a box around the halo and then discarding particles which are not associated with the halo.

\begin{lstlisting}[style=code, language=Python]
import numpy as np
import unyt
import hdfstream
import swiftsimio as sw

# Connect to the hdfstream service and open the root directory
root_dir = hdfstream.open("cosma", "/")

# Open the z=0 halo catalogue from the L1_m10_DMO simulation
soap_file = root_dir["FLAMINGO/L1_m10/L1_m10_DMO/SOAP-HBT/halo_properties_0077.hdf5"]
soap = sw.load(soap_file)

# Get halo positions, masses and indexes
halo_pos = soap.input_halos.halo_centre
halo_m200c = soap.spherical_overdensity_200_crit.total_mass
halo_index = soap.input_halos.halo_catalogue_index

# Select the most massive halo
i = np.argmax(halo_m200c)
target_pos = halo_pos[i,:]
target_index = halo_index[i]

# Choose a region to read in. Note that we need to choose a radius large
# enough to enclose all halo particles.
radius = 5.0*unyt.Mpc
region = [[x-radius, x+radius] for x in target_pos]

# Open the z=0 snapshot from the L1_m10_DMO simulation and select this region
snap_file = root_dir["FLAMINGO/L1_m10/L1_m10_DMO/snapshots/flamingo_0077/flamingo_0077.hdf5"]
mask = sw.mask(snap_file)
mask.constrain_spatial(region)
snap = sw.load(snap_file, mask=mask)

# Read position and halo index of dark matter particles in this region
dm_pos = snap.dark_matter.coordinates
dm_halo_index = snap.dark_matter.halo_catalogue_index

# Of the particles we read, identify those which belong to the halo
in_halo = (dm_halo_index == target_index)
halo_dm_pos = dm_pos[in_halo,:]
\end{lstlisting}
The result is an array containing the positions of the particles in the halo. Other particle properties can be read in a similar way.